\documentstyle[12pt,epsfig]{article}
\begin{document}

\begin{center}
{\large {\bf On the Alpha Activity of Natural Tungsten Isotopes}}
\end{center}

\vskip 0.2cm

\begin{center}
F.A.~Danevich$^a$, A.Sh.~Georgadze$^a$, V.V.~Kobychev$^{a,}$\footnote{%
Current address: INFN, Laboratori Nazionali del Gran Sasso, 67010 Assergi
(AQ), Italy}, S.S.~Nagorny$^a$, A.S.~Nikolaiko$^a$, O.A.~Ponkratenko$^a$,
V.I.~Tretyak$^a$, S.Yu.~Zdesenko$^a$, Yu.G.~Zdesenko$^{a,}$\footnote{%
Corresponding author. {\it E-mail address:} zdesenko@kinr.kiev.ua}

\noindent $^a${\it Institute for Nuclear Research, MSP 03680 Kiev, Ukraine}

P.G.~Bizzeti$^b$, T.F.~Fazzini$^b$, P.R.~Maurenzig$^b$

\noindent $^b${\it Dipartimento di Fisica, Universit$\acute a$ di Firenze
and INFN, I-50019 Firenze, Italy}

\hspace{1.0in}

{\bf Abstract}
\end{center}

\noindent
The indication for the $\alpha $ decay of $^{180}$W with a half-life $%
T_{1/2}^\alpha $=1$.1_{-0.4}^{+0.8}$(stat)$\pm 0.3$(syst)$\times $1$0^{18}$
yr has been observed for the first time with the help of the super-low
background $^{116}$CdWO$_4$ crystal scintillators. In conservative approach
the lower limit on half-life of $^{180}$W has been established as $%
T_{1/2}^\alpha (^{180}$W) $\geq $ 0.7$\times 10^{18}$ yr at 90\% C.L.
Besides, new $T_{1/2}^\alpha $ bounds were set for $\alpha $ decay of $%
^{182} $W, $^{183}$W, $^{184}$W and $^{186}$W at the level of 10$^{20}$~yr.

\vskip 0.3cm

\noindent PACS number(s): 23.60.+e; 29.40.Mc

\vskip 0.3cm


\section{INTRODUCTION}

Alpha decay is allowed energetically for the five naturally occurring
isotopes of tungsten \cite{Aud95}, but it was never observed up to now. The $%
\alpha $ activity of tungsten with $\alpha $ particle energy of about 3 MeV
and half-life $T_{1/2}=$ 2.2$\times \delta \times $1$0^{17}$ yr (where $%
\delta $ is the relative abundance of isotope) was declared in an early
experiment with nuclear emulsion technique \cite{Por53}. Because $Q_\alpha $
value for all natural tungsten nuclides, except $^{180}$W, is less than 2
MeV, the result of ref. \cite{Por53} could be attributed to $\alpha $ decay
of $^{180}$W ($\delta \approx 0.12\%$ \cite{abundance}) with $T_{1/2}$ = 2.6$%
\times $1$0^{14}$ yr. However, this observation was ruled out in the work
\cite{Bea60}, where the cadmium tungstate (CdWO$_4$) crystal scintillator
(mass of 20.9 g) was used as a source and detector of $\alpha $ decay events
simultaneously. After 193 h of measurements the limit $T_{1/2}$ $\geq $ 1.0$%
\times $1$0^{15}$ yr was established \cite{Bea60}. A similar restriction ($%
T_{1/2}$ $\geq $ 9.0$\times $10$^{14}$ yr) was also obtained in experiment
(66.7 h of exposition) with ionization counter (1200 cm$^2$ in area) and
thin (83 $\mu $g/cm$^2$) sample of W (with total mass of 79 mg) enriched in $%
^{180}$W to 6.95\% \cite{Mac61}.

These bounds were improved only recently in the measurements with two
scintillators: CdWO$_4$ (mass of 452 g, running time of 433 h), and $^{116}$%
CdWO$_4$ enriched in $^{116}$Cd to 83\% (91.5 g, 951 h) \cite{Geo95}. The
limits on the half-lives for $\alpha $ decay of different W isotopes were
set in the range of $\approx $ 10$^{17}$--1$0^{19}$ yr (see Table).

\begin{table}[!htbp]
\begin{center}
\caption{Theoretical and experimental half-lives (or limits at 90\%
C.L.) for $\alpha $ decay of natural W isotopes obtained in present work.
The uncertainties of calculated $T_{1/2}$ values are related with
uncertainties of the $Q_\alpha $. The most stringent previous experimental
bounds \cite{Geo95} are cited for comparison.}
\begin{tabular}{|c|ccccc|}
\hline
Isotope, & \multicolumn{1}{|c|}{$Q_\alpha $, MeV} & \qquad Calculated &
\multicolumn{1}{c|}{$T_{1/2}$, yr\qquad \quad} & Experimental & $T_{1/2}$,
yr\quad \\ \cline{3-6}\cline{3-4}
abundance & \multicolumn{1}{|c|}{\cite{Aud95}} & \multicolumn{1}{c|}{based
on \cite{Buc91}} & \multicolumn{1}{c|}{based on \cite{Poe83}} &
\multicolumn{1}{c|}{present work} & \cite{Geo95} \\
\cite{abundance} & \multicolumn{1}{|c|}{} & \multicolumn{1}{c|}{} &
\multicolumn{1}{c|}{} & \multicolumn{1}{c|}{} &  \\ \hline
$^{180}$W & \multicolumn{1}{|c|}{2.516(5)} & \multicolumn{1}{c|}{$%
8.3_{-1.3}^{+1.6}\times $1$0^{17}$} & \multicolumn{1}{c|}{$%
2.0_{-0.3}^{+0.4}\times $1$0^{18}$} & \multicolumn{1}{c|}{$%
1.1_{-0.5}^{+0.9}\times $1$0^{18}$} &  \\
0.12(1)\% & \multicolumn{1}{|c|}{} & \multicolumn{1}{c|}{} &
\multicolumn{1}{c|}{} & \multicolumn{1}{c|}{$\geq 0$.7$\times 10^{18}$} & $%
\geq 7.$4$\times $1$0^{16}$ \\ \hline
$^{182}$W & \multicolumn{1}{|c|}{1.774(3)} & \multicolumn{1}{c|}{$%
3.0_{-0.5}^{+0.5}\times $1$0^{32}$} & \multicolumn{1}{c|}{$%
1.4_{-0.2}^{+0.3}\times $1$0^{33}$} & \multicolumn{1}{c|}{$\geq 1.$7$\times
10^{20}$} & $\geq 8.$3$\times 10^{18}$ \\
26.50(16)\% & \multicolumn{1}{|c|}{} & \multicolumn{1}{c|}{} &
\multicolumn{1}{c|}{} & \multicolumn{1}{c|}{} &  \\ \hline
$^{183}$W & \multicolumn{1}{|c|}{1.682(3)} & \multicolumn{1}{c|}{--} &
\multicolumn{1}{c|}{$>5.7_{-1.0}^{+1.3}\times $1$0^{38}$} &
\multicolumn{1}{c|}{$\geq 0.$8$\times $1$0^{20}$} & $\geq 1.$9$\times $1$%
0^{18}$ \\
14.31(4)\% & \multicolumn{1}{|c|}{} & \multicolumn{1}{c|}{} &
\multicolumn{1}{c|}{} & \multicolumn{1}{c|}{} &  \\ \hline
$^{184}$W & \multicolumn{1}{|c|}{1.659(3)} & \multicolumn{1}{c|}{$%
3.8_{-0.6}^{+0.9}\times $1$0^{35}$} & \multicolumn{1}{c|}{$%
2.6_{-0.5}^{+0.6}\times $1$0^{36}$} & \multicolumn{1}{c|}{$\geq 1.$8$\times $%
1$0^{20}$} & $\geq 4.$0$\times $1$0^{18}$ \\
30.64(2)\% & \multicolumn{1}{|c|}{} & \multicolumn{1}{c|}{} &
\multicolumn{1}{c|}{} & \multicolumn{1}{c|}{} &  \\ \hline
$^{186}$W & \multicolumn{1}{|c|}{1.123(7)} & \multicolumn{1}{c|}{$%
8.7_{-4.9}^{+11.2}\times $1$0^{55}$} & \multicolumn{1}{c|}{$%
2.0_{-1.1}^{+2.6}\times $1$0^{57}$} & \multicolumn{1}{c|}{$\geq 1.$7$\times
10^{20}$} & $\geq 6.$5$\times $1$0^{18}$ \\
28.43(19)\% & \multicolumn{1}{|c|}{} & \multicolumn{1}{c|}{} &
\multicolumn{1}{c|}{} & \multicolumn{1}{c|}{} &  \\ \hline
\end{tabular}
\end{center}
\end{table}

In this paper the new results of Kiev-Firenze experiment (2975 h of
exposition) on $\alpha $ decay of natural tungsten isotopes are described
(the preliminary analysis of the 975 h data was presented in \cite{Sarov}).
They were obtained with the help of the super-low background spectrometer
\cite{Dan00} based on enriched $^{116}$CdWO$_4$ crystal scintillators. The
sensitivity of this apparatus to measure $\alpha $ activity was enhanced
substantially in comparison with our previous work \cite{Geo95} mainly due
to the developed technique of pulse-shape analysis of the data, which allows
us to distinguish events caused by $\alpha $ particles from those by $\gamma
$ rays ($\beta $ particles).

\section{EXPERIMENT AND DATA ANALYSIS}

\subsection{Experimental set-up}

The set-up, used in our study, was originally devoted to the search for the
neutrinoless double beta decay of $^{116}$Cd \cite{Dan00}. The 2$\beta $
decay experiment is carried out by the INR (Kiev)\footnote{%
From 1998 by the Kiev-Firenze collaboration \cite{Dan00}.} in the Solotvina
Underground Laboratory (allocated in a salt mine 430 m underground \cite
{Zde87}) and results were published elsewhere \cite{Dan00,Dan95}. The high
purity $^{116}$CdWO$_4$ crystal scintillators, enriched in $^{116}$Cd to
83\%, were developed and grown for the search \cite{Dan95}. Their light
output is $\approx $30--35\% of NaI(Tl). The fluorescence peak emission is
at 480 nm with principal decay time of $\approx $14 $\mu $s \cite{Faz98}.
The CdWO$_4$ refractive index equals 2.3. The density of crystal is 7.9 g/cm$%
^3$, and material is non-hygroscopic and chemically inert. In the apparatus
(see for details \cite{Dan00}) four $^{116}$CdWO$_4$ crystals (total mass
330 g) are exploited. They are viewed by a low background 5$^{^{\prime
\prime }}$ EMI phototube (PMT) with RbCs photocathode through light-guide $%
\oslash $10$\times $55 cm, which is glued of two parts: quartz (25 cm) and
plastic scintillator (Bicron BC-412, 30 cm). The enriched $^{116}$CdWO$_4$
crystals are surrounded by an active shield made of 15 natural CdWO$_4$
crystals of large volume with total mass of 20.6 kg. These are viewed by a
PMT through an active plastic light-guide $\oslash $17$\times $49 cm. The
whole CdWO$_4$ array is situated within an additional active shield made of
plastic scintillator 40$\times $40$\times $95 cm, thus, together with both
active light-guides, a complete 4$\pi $ active shield of the main ($^{116}$%
CdWO$_4$) detector is provided.

The outer passive shield consists of high purity copper ($3$ -- $6$ cm),
lead ($22.5$ -- $30$ cm) and polyethylene (16 cm). Two plastic scintillators
(120$\times $130$\times $3 cm) are installed above the passive shield and
are used as cosmic muon veto counters. The set-up is carefully isolated
against environment radon penetration. All materials for the installation
were previously tested and selected for low radioactive impurities in order
to reduce their contributions to background.

An event-by-event data acquisition system records the amplitude, arrival
time, additional tags (the coincidence between different detectors) and
pulse shape (in 2048 channels with 50 ns channel's width) of the $^{116}$CdWO%
$_4$ detector in the energy range of $0.08-5$ MeV.

The energy scale and resolution of the spectrometer were determined with the
$\gamma $ sources $^{22}$Na, $^{40}$K, $^{60}$Co, $^{137}$Cs, $^{207}$Bi, $%
^{226}$Ra, $^{232}$Th and $^{241}$Am. The energy dependence of the
resolution in the energy interval $60-2615$ keV is expressed as follows: FWHM%
$_\gamma = -44 + \sqrt{2800+23.4\cdot E_\gamma},$ where
energy $E_\gamma $ and FWHM$_\gamma $ are in keV. For instance, energy
resolutions of 33.7\%, 13.5\%, 11.5\% and 8.0\% were measured for $\gamma $
lines with the energies of 60, 662, 1064 and 2615 keV, respectively. The
routine calibration was carried out with a $^{207}$Bi (weekly) and $^{232}$%
Th (monthly) $\gamma $ sources. The dead time of the spectrometer and data
acquisition was permanently controlled with the help of a light emitting
diode optically connected to the main PMT (typical value was about 14\%).

Due to active and passive shields, and as a result of the time-amplitude
\cite{Dan00} and pulse-shape analysis of the data \cite{Faz98}, the
background rate of $^{116}$CdWO$_4$ detectors in the energy region $2.5$ -- $%
3.2$ MeV (0$\nu $2$\beta $ decay energy of $^{116}$Cd is 2.8 MeV) was
reduced to 0.04 counts/yr$\cdot $kg$\cdot $keV. It is the lowest background
which has ever been reached with crystal scintillators.

\subsection{Response of the $^{116}$CdWO$_{4}$ detector to $\alpha $
particles}

It is well known that relative ratio of the CdWO$_4$ scintillation light
output for $\alpha $ and $\beta $ particles with the same energies
(so-called $\alpha /\beta $ ratio\footnote{%
The detector energy scale is measured with $\gamma $ sources, thus the
notation ``$\alpha /\gamma $ ratio'' could be more adequate. However,
because $\gamma $ rays interact with matter by means of the energy transfer
to electrons, in present paper we are using traditional notation ``$\alpha
/\beta $ ratio''.}) depends on the energy of the absorbed particles \cite
{Geo95}. Because we are looking for the $\alpha $ decay of W isotopes, such
a dependence must be precisely measured in the $\alpha $ energy range of
1--3 MeV (see Table). Unfortunately, among nuclides from U/Th families there
are no $\alpha $ emitters with such energies (the lowest available $\alpha $
energy is 4.0 MeV from $^{232}$Th). To overcome this problem, a special
method of calibration was developed, in which a collimated beam of $\alpha $
particles from $^{241}$Am source was passed through a thin absorber with
known thickness, thus the energy of $\alpha $ particles after absorber can
be calculated precisely. Furthermore, it was measured with the help of a
surface-barrier semiconductor detector. The dimensions of the collimator are
$\oslash $0.75 $\times $~2 mm, and thickness of a single mylar film absorber
is 0.65 mg/cm$^2.$ By using this technique and different sets of absorbers, $%
\alpha $ particles with energies of 0.46, 2.07, 3.04, 3.88, 4.58, and 5.25
MeV were obtained, which allows us to calibrate our detector in the energy
range of interest.

It is also known that the light output of crystal scintillators may depend
on the direction of $\alpha $ irradiation relative to the crystal axes \cite
{Birks}. To study this effect for CdWO$_4$, $^{116}$CdWO$_4$ and CaWO$_4$
crystals, they were irradiated by $\alpha $ particles in three directions
perpendicular to the (010), (001) and (100) crystal planes\footnote{%
In the $^{116}$CdWO$_4$ crystal (010) plane is perpendicular to the cylinder
axis, so for the (010) direction $\alpha $ source was placed on the top of
crystal (in the center of the flat circle surface). For the (001) and (100)
directions crystal was irradiated in the middle of cylindrical surface.},
and three experimental dependences of the $\alpha /\beta $ ratio
(corresponding to three mentioned directions) as a function of $\alpha $
particle energy were derived from measurements. However, in a real crystal
the amplitude of a light signal depends also on the point, from which
scintillation light is emitted (so-called nonuniformity of light
collection). The latter can distort the anticipated effect of crystal's
orientation, and hence, should be properly taken into account. With this
aim, the light propagation in the CaWO$_4$ scintillator, for which the
effect of crystal orientation was not observed, was simulated with the help
of the GEANT3.21 package \cite{GEANT} for the light emitting points on the
top and side surfaces of the crystals (with dimensions 20$\times $20$\times $%
11 and 40$\times $34$\times $23 mm), as well as for those uniformly
distributed inside the crystal. It was found that results of simulations are
in a good agreement with experiment. Then, the same simulations were
performed for the $^{116}$CdWO$_4$ crystal ($\oslash 32\times 19$ mm), and
results of each calibration measurement with this crystal were corrected by
using simulated distributions of the light collection for $\alpha $
particles emitted from the corresponding point. So, values of the $\alpha
/\beta $ ratio for the direction 1 was multiplied by a factor of 0.85, while
for the direction 2 and 3 -- by a factor of 1.08. As an example, such
corrected dependences of the $\alpha /\beta $ ratio versus the energy and
direction of $\alpha $ particles are shown in Fig. 1 for enriched $^{116}$%
CdWO$_4$ crystal ($\oslash 32\times 19$ mm).

\nopagebreak
\begin{figure}[ht]
\begin{center}
\mbox{\epsfig{figure=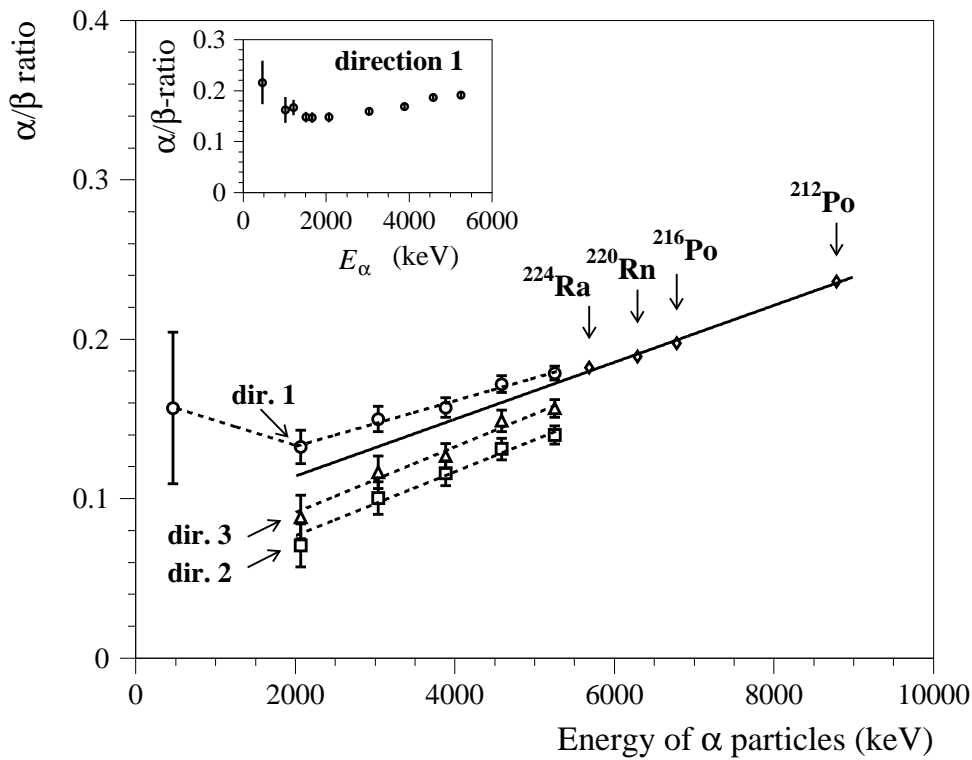,height=10.0cm}}
\caption{The dependence of the $\alpha /\beta $ ratio on energy and
direction of incident $\alpha $ particles measured with the enriched $^{116}$%
CdWO$_4$ scintillator ($\oslash 32\times 19$ mm). The crystal was irradiated
by external $\alpha $ source in directions perpendicular to (010), (001) and
(100) crystal planes (dir. 1, 2 and 3, respectively). Besides, internal $%
\alpha $ peaks of $^{224}$Ra, $^{220}$Rn, $^{216}$Po and $^{212}$Po from the
intrinsic contamination of the $^{116}$CdWO$_4$ crystal were used (see
text). Solid lines represent the fit of $\alpha $/$\beta $ ratio dependence.
In the insert the behaviour of $\alpha /\beta $ ratio measured with CdWO$_4$
scintillator ($\oslash 25\times 0.9$ mm) in the direction 1 is shown.}
\end{center}
\end{figure}

In addition to the measurements with the external source, the $\alpha $
peaks from the intrinsic trace contamination of the $^{116}$CdWO$_4$
crystals by nuclides from Th chain were also used for determination of the $%
\alpha /\beta $ ratio. These peaks were selected from the background by
using the time-amplitude analysis \cite{Dan00,GSO}. For instance, the
following sequence of $\alpha $ decays from the $^{232}$Th family was
searched for and observed: $^{224}$Ra ($Q_\alpha $ = $5.79$ MeV, $T_{1/2}$ =
$3.66$ d) $\rightarrow $ $^{220}$Rn ($Q_\alpha $ = $6.40$ MeV, $T_{1/2}$ = $%
55.6$ s) $\rightarrow $ $^{216}$Po ($Q_\alpha $ = $6.91$ MeV, $T_{1/2}$ = $%
0.145$ s) $\rightarrow $ $^{212}$Pb. The obtained $\alpha $ peaks ($\alpha $
nature of events was confirmed by a pulse-shape analysis described below),
as well as the distributions of the time intervals between events are in a
good agreement with those expected for $\alpha $ particles of $^{224}$Ra, $%
^{220}$Rn and $^{216}$Po (see Fig. 2). On this basis the activity of $^{228}$%
Th in $^{116}$CdWO$_4$ crystals was determined as 39(2) $\mu $Bq/kg\footnote{%
The same technique was applied to the sequence of decays from the $^{235}$U
and $^{238}$U families. Activity of 5.5(14) $\mu $Bq/kg for $^{227}$Ac ($%
^{235}$U family) and limit $\leq $ 5 $\mu $Bq/kg for $^{226}$Ra ($^{238}$U
family) in the $^{116}$CdWO$_4$ crystals were set.}. Moreover, the $\alpha $
peak of $^{212}$Po (the daughter of the $^{212}$Bi) was reconstructed with
the help of the front edge analysis of the scintillation signals. The energy
and time distributions of the sequence of $\beta $ decay of $^{212}$Bi and $%
\alpha $ decay of $^{212}$Po, selected from the background, are presented in
Fig. 3 (a, b and c), while typical example of such an event is shown in Fig.
3d.

\nopagebreak
\begin{figure}[ht]
\begin{center}
\mbox{\epsfig{figure=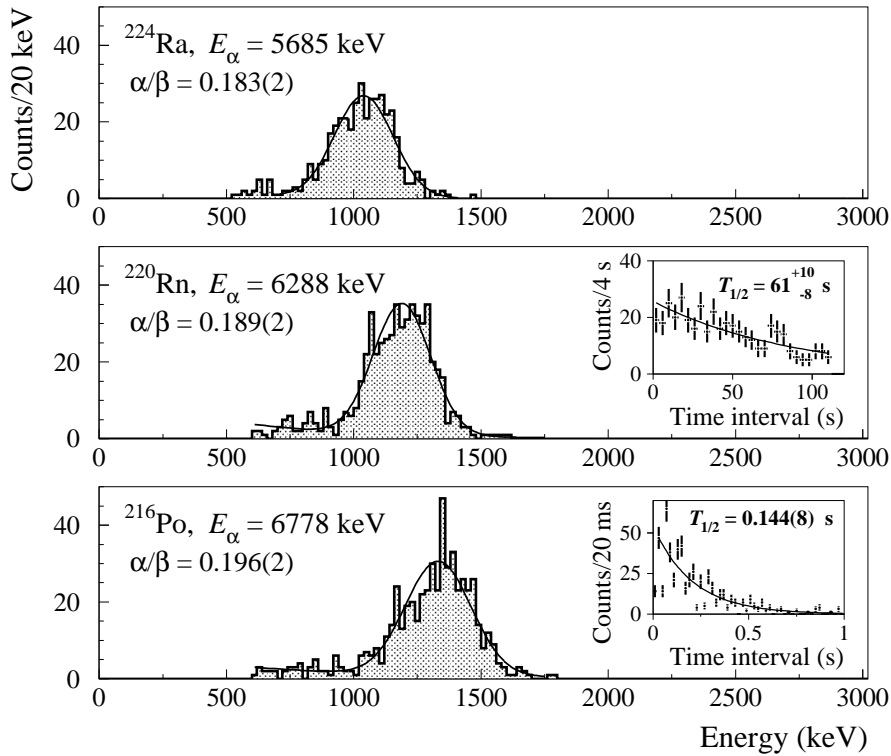,height=10.0cm}}
\caption{The $\alpha $ peaks of $^{224}$Ra, $^{220}$Rn, $^{216}$Po
selected by the time-amplitude analysis from background data accumulated
during 14745 h with the $^{116}$CdWO$_4$ detector. In the inserts the time
distributions between the first and second (and between second and third)
events together with exponential fits are presented. Obtained half-lives of $%
^{220}$Rn and $^{216}$Po ($61_{-8}^{+10}$ s and $0.144(8)$ s, respectively)
are in a good agreement with the table values: $55.6(1)$ s and $0.145(2)$ s
\cite{Fir96}.}
\end{center}
\end{figure}

\nopagebreak
\begin{figure}[ht]
\begin{center}
\mbox{\epsfig{figure=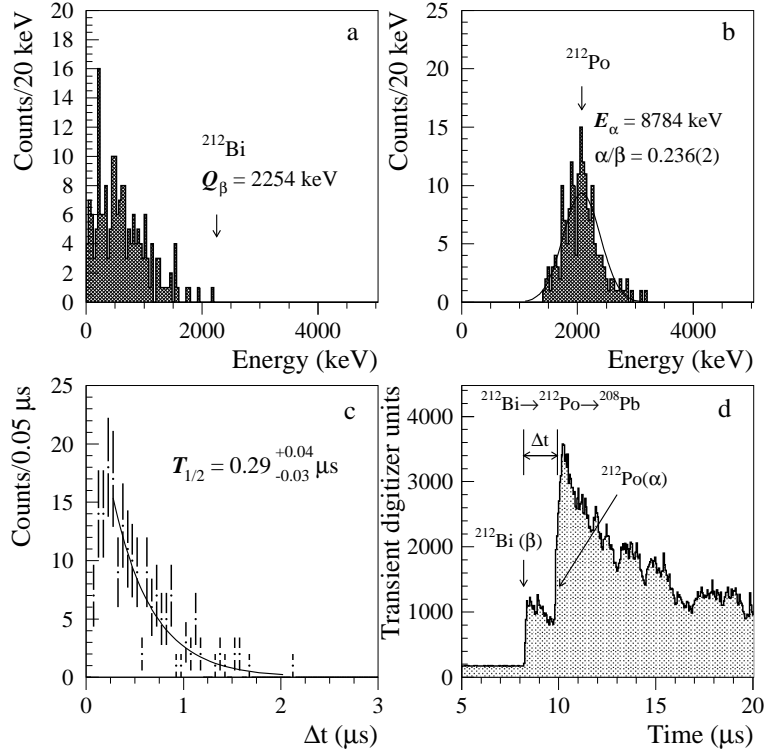,height=10.0cm}}
\caption{The energy (a, b) and time (c) distributions for fast
sequence of $\beta $ ($^{212}$Bi, $Q_\beta =2254$ keV) and $\alpha $ ($%
^{212} $Po, $E_\alpha =8784$ keV, $T_{1/2}=0.299(2)~\mu $s) decays selected
from the background data by the pulse-shape analysis. (d) Example of such an
event in the $^{116}$CdWO$_4$ scintillator.}
\end{center}
\end{figure}

The values of the $\alpha $/$\beta $ ratio, derived with the internal $%
\alpha $ peaks, and their fit are depicted in Fig. 1. Fit yields for the
energy range $2.0-8.8$ MeV: $\alpha /\beta =0.083(9)+0.0168(13)\cdot
E_\alpha $, where $E_\alpha $ is in MeV.\footnote{%
The growth of the $\alpha $/$\beta $ ratio with $\alpha $ particles energy
has been earlier observed for cadmium tungstate \cite{Geo95} and other
scintillators \cite{Birks,GSO,Barton}.} Because in measurements with
internal sources the effect of crystal's orientation is averaged, such an
extrapolation of the fit into the energy region 2.0 -- 5.5 MeV is reasonable
and it was also proved by the behaviour of the $\alpha $/$\beta $ ratios
measured with external sources (see Fig. 1). At the same time, in the $%
\alpha $ energy interval of $0.5-2.0$ MeV we find that $\alpha $/$\beta $
ratio is decreased with the energy. It was also confirmed by the measurement
with the thin ($\oslash 25\times 0.9$ mm) CdWO$_4$ crystal scintillator (see
insert in Fig. 1). Similar behaviour (fall with energy in the $10-100$ keV
energy region) of relative scintillation efficiency for Ca and F recoil
nuclei in CaF$_2$(Eu) scintillator was reported in \cite{Tovey}. Thus, for $%
0.5-2.0$ MeV energy range we obtain: $\alpha /\beta =0.15(3)-0.015(8)\cdot
E_\alpha ,$ where $E_\alpha $ is in MeV.

The calibration data were also used to determine the energy resolution of
the detector for $\alpha $ particles: FWHM$_\alpha $(keV)$=33+0.247E_\alpha
^\gamma $, where $E_\alpha ^\gamma $ is the energy of $\alpha $ particles in
$\gamma $ scale expressed in keV.

\subsection{Pulse-shape analysis}

The pulse-shape analysis of CdWO$_4$ scintillation signals was developed on
the basis of the optimal digital filter \cite{Gat62}, and clear
discrimination between $\gamma $ rays (electrons\footnote{%
Because $\gamma $ rays interact with matter by means of the energy transfer
to electron, it was assumed and experimentally proved with $\beta $
particles from the decay of internal $^{113}$Cd that pulse shapes for
electrons and $\gamma $'s are the same.}) and $\alpha $ particles was
achieved \cite{Faz98}. The pulse shape of cadmium tungstate scintillators
can be described as: $f(t)=\sum A_i/(\tau _i-\tau _0)\cdot (e^{-t/\tau
_i}-e^{-t/\tau _0})$, where $A_i$ are amplitudes and $\tau _i$ are decay
constants for different light emission components, $\tau _0$ is the
integration constant of electronics ($\approx 0.18~\mu $s). To provide an
analytic description of the $\alpha $ or $\gamma $ signals, the pulse shape
resulting from the average of a large number of individual events has been
fitted with the sum of three (for $\alpha $ particles) or two (for $\gamma $%
-s ) exponents, giving the reference pulse shapes $\overline{f}_\alpha (t)$
and $\overline{f}_\gamma (t)$ (see for details ref. \cite{Faz98}). For the
enriched crystals (used in the experiment) the following values were
obtained: $A_1^\alpha $=80.9, $\tau _1^\alpha $=12.7~$\mu $s, $A_2^\alpha $%
=13.4, $\tau _2^\alpha $=3.3 $\mu $s, $A_3^\alpha $=5.7, $\tau _3^\alpha $%
=0.96 $\mu $s for $\approx $5 MeV $\alpha $ particles and $A_1^\gamma $%
=94.2, $\tau _1^\gamma $=13.6 $\mu $s, $A_2^\gamma $=5.6, $\tau _2^\gamma $%
=2.1 $\mu $s for $\approx $1 MeV $\gamma $ quanta.

In the data processing the digital filter was applied to each experimental
signal $f(t)$ with aim to obtain the numerical characteristic of its shape
defined as $SI$ (shape indicator): $SI=\sum f(t_k)\times P(t_k)/\sum f(t_k)$%
, where the sum is over time channels $k,$ starting from the origin of
signal and up to 50 $\mu $s, $f(t_k)$ is the digitized amplitude (at the
time $t_k$) of a given signal. The weight function $P(t)$ is defined as:

\begin{center}
$P(t)=\{\overline{f}_\alpha (t)-\overline{f}_\gamma (t)\}/\{\overline{f}%
_\alpha (t)+\overline{f}_\gamma (t)\}$.
\end{center}

The $SI$ distributions measured with different $\alpha $ and $\gamma $
sources (see some examples in Fig. 4) are well described by Gaussian
functions, whose standard deviations $\sigma (SI_\alpha )$ and $\sigma
(SI_\gamma )$ depend on energy: $\sigma (SI_\alpha $) =--0$.$3$+$0.7$\times $%
1$0^{-3}E_\alpha ^\gamma +$1$700/E_\alpha ^\gamma $ for $\alpha $ particles,
and $\sigma (SI_\gamma )=1.5$1$-$0$.2$3$\times $1$0^{-3}E_\gamma +$4$%
02/E_\gamma $ for $\gamma $ quanta, where $E_\alpha ^\gamma $ and $E_\gamma $
are in keV. As it is seen from Fig. 4, electrons ($\gamma $ rays) and $%
\alpha $ particles are clearly distinguished for the energies above 0.6 MeV (%
$E_\alpha $ $\approx $3 MeV), while pulse shape discrimination ability of
CdWO$_4$ scintillator is worsening at lower energies. Nevertheless, it is
also visible from Fig. 4a that even 2 MeV $\alpha $ particles ($E_\alpha
^\gamma $ $\approx $0.3 MeV) can be separated from $\gamma $ background with
reasonable accuracy. For instance, with requirement that $\approx $87\% of $%
\alpha $ particles must be registered we will get about 13\% of $\gamma $
events as background (see Fig. 4a).

\nopagebreak
\begin{figure}[ht]
\begin{center}
\mbox{\epsfig{figure=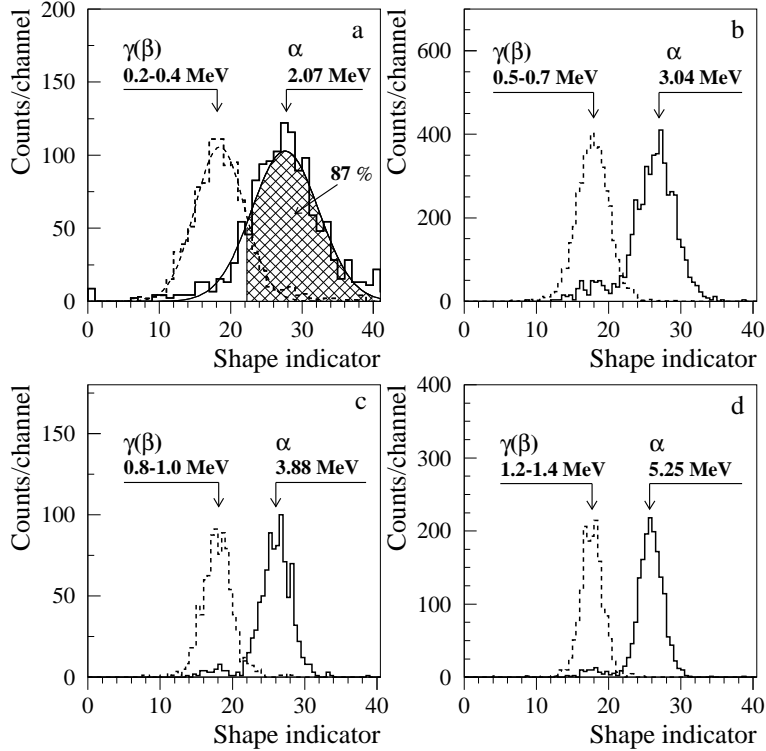,height=10.0cm}}
\caption{The examples of the shape indicator distributions measured
by the enriched $^{116}$CdWO$_4$ crystal scintillator ($\oslash $32$\times $%
19 mm) with $\alpha $ particles and $\gamma $ rays with the different
energies: (a) $E_\gamma $=~0.2--0.4 MeV, $E_\alpha $ = 2.07 MeV; (b)
$E_\gamma $= 0.5--0.7 MeV, $E_\alpha $ = 3.04 MeV; (c) $E_\gamma $= 0.8--1.0
MeV, $E_\alpha $ = 3.88 MeV; (d) $E_\gamma $= 1.2--1.4 MeV, $E_\alpha $ =
5.25 MeV.}
\end{center}
\end{figure}

The dependence of the pulse shape on energy and direction of $\alpha $
particles measured with the enriched $^{116}$CdWO$_4$ crystal scintillator ($%
\oslash 32\times $19 mm) is presented in Fig.~5. In the energy range of $%
0.5-7.0$ MeV the average dependence of the shape indicator on the energy can
be approximated by the function: $SI_\alpha =29.5-0.19$5$\times $1$%
0^{-2}E_\alpha ^\gamma $, where $E_\alpha ^\gamma $ is in keV.

\nopagebreak
\begin{figure}[ht]
\begin{center}
\mbox{\epsfig{figure=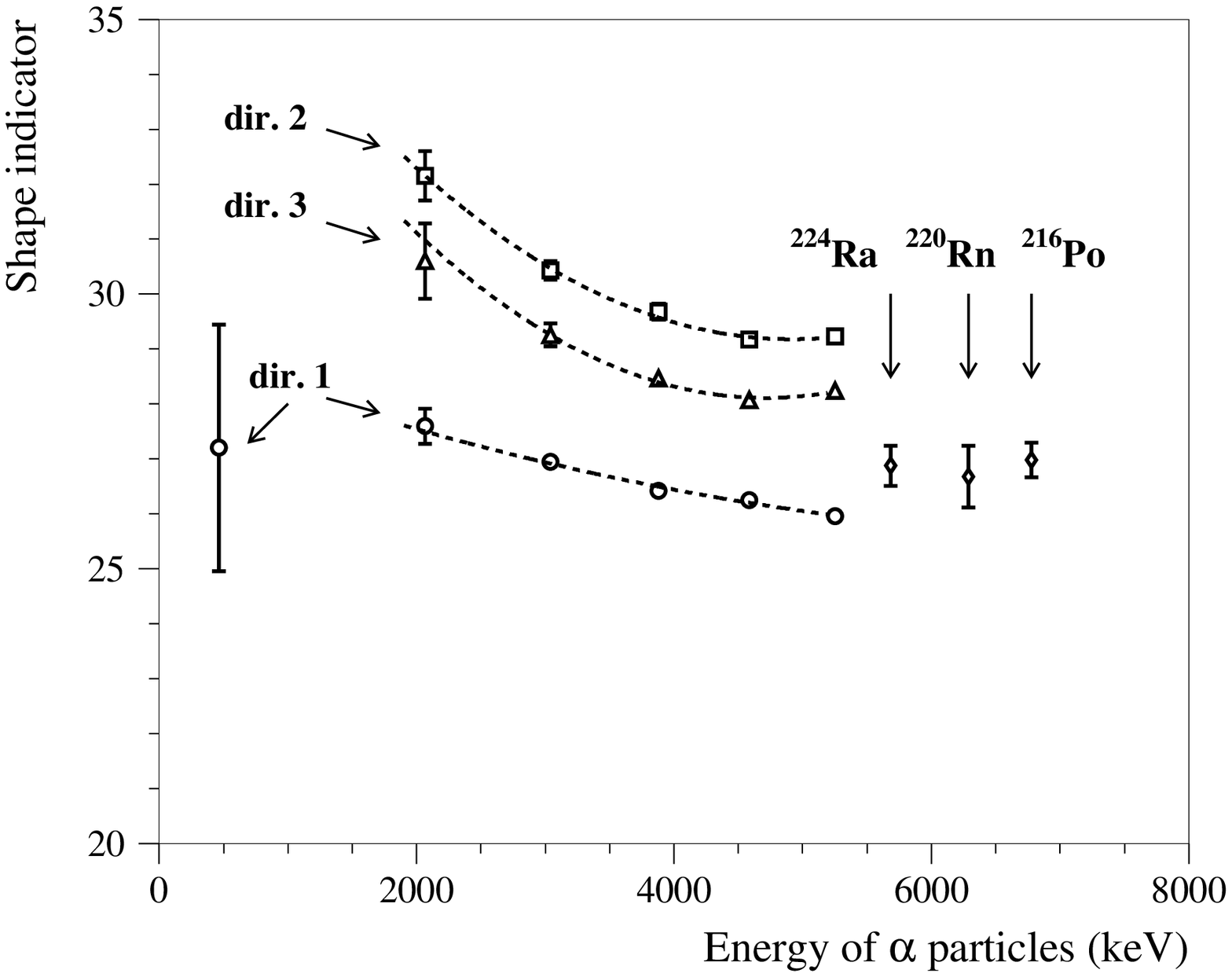,height=10.0cm}}
\caption{The dependence of the shape indicator on the energy and the
direction of incident $\alpha $ particles measured with the enriched $^{116}$%
CdWO$_4$ crystal scintillator. The $\alpha $ peaks of $^{224}$Ra, $^{220}$Rn
and $^{216}$Po were selected by the time-amplitude analysis of background
data.}
\end{center}
\end{figure}

For $\gamma $ quanta the energy dependence of the shape indicator was
measured with $\gamma $ sources in the energy range of $0.04-3.2$ MeV: $%
SI_\gamma =18.4-0.11$7$\times $1$0^{-2}E_\gamma +0.5$4$\times $1$%
0^{-6}E_\gamma ^2$, where $E_\gamma $ is in keV.

Besides, a digital filter for the pulses of the plastic scintillator light
guide was developed and clear separation was obtained for the pure events in
the plastic and cadmium tungstate scintillators. It allows us to
discriminate plastic pulses in more complicated cases, when they are
overlapped with the signals of $^{116}$CdWO$_4$ crystals. The use of this
filter leads to some loss of the registration efficiency for the events in
the $^{116}$CdWO$_4$ detector (of the order of $\approx $5\%), which,
however, can be correctly taken into account on the basis of calibration
measurements.

\section{RESULTS AND DISCUSSION}

\subsection{Background interpretation}

The background spectra of ($\gamma +\beta $) and $\alpha $ events measured
by four $^{116}$CdWO$_4$ crystals (330 g, exposition 2975 h) are depicted in
Fig. 6.

\nopagebreak
\begin{figure}[ht]
\begin{center}
\mbox{\epsfig{figure=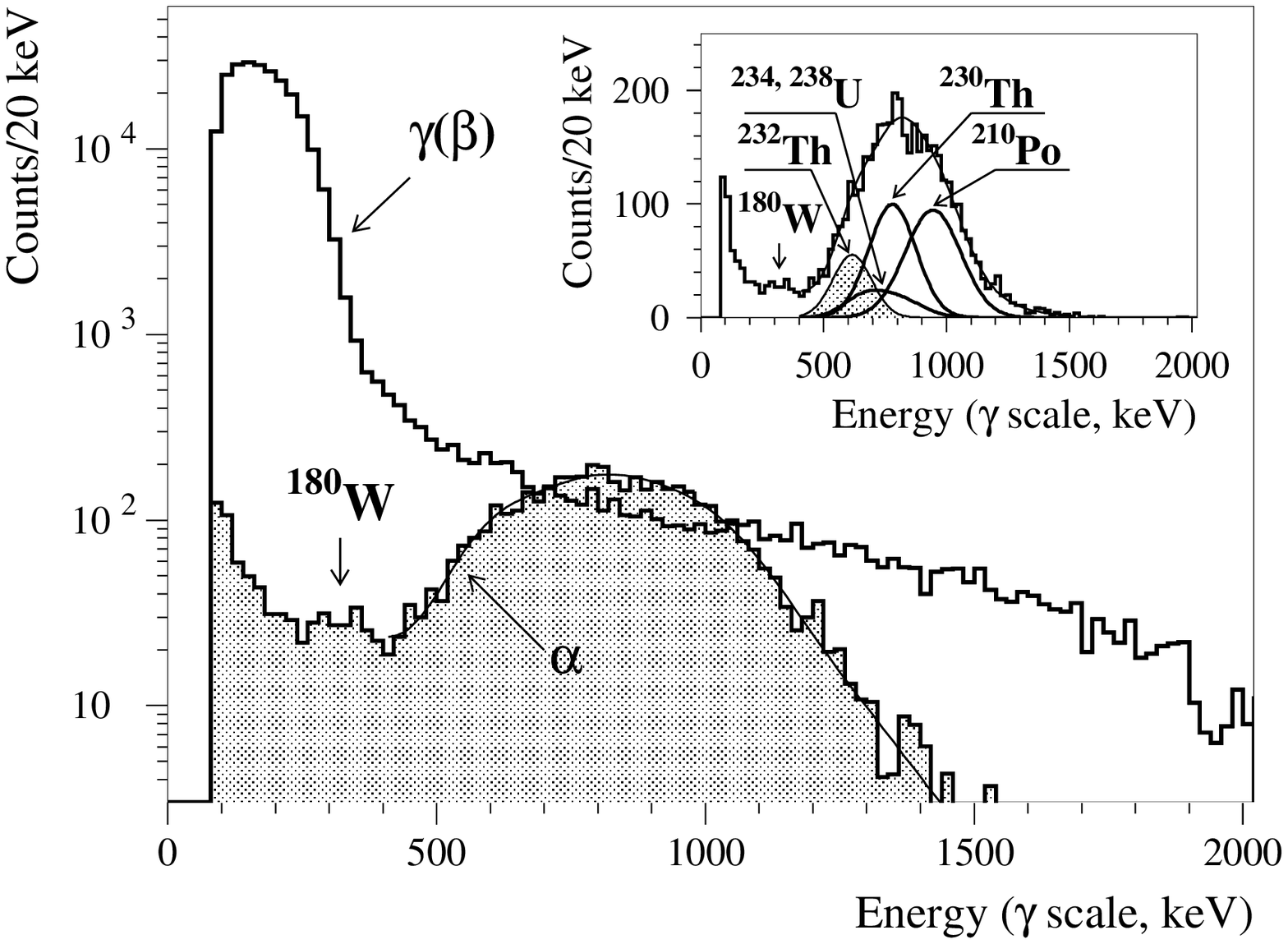,height=10.0cm}}
\caption{Energy distributions of $\gamma (\beta )$ and $\alpha $
events, which were selected by the pulse shape analysis from the data of $%
^{116}$CdWO$_4$ crystals (330 g) measured during 2975 h. In the insert, $%
\alpha $ spectrum is depicted together with the model, which includes $%
\alpha $ decays from $^{232}$Th and $^{238}$U families. The total $\alpha $
activity in the $^{116}$CdWO$_4$ crystals is 1.40(10) mBq/kg.}
\end{center}
\end{figure}

The $\gamma (\beta )$ spectrum shown in Fig. 6 was built by selecting the
following interval of $SI$ values: $SI_\gamma -2.4\sigma _\gamma
<SI<SI_\gamma +2.4\sigma _\gamma $, which contains 98\% of $\gamma (\beta )$
events. In the low energy region the background of the $^{116}$CdWO$_4$
detector is caused mainly by the fourth-forbidden $\beta $ decay of $^{113}$%
Cd ($T_{1/2}=7.7\times 10^{15}$ yr \cite{Cd113}, Q$_\beta =316$ keV \cite
{Fir96}) which is present in the enriched crystal with abundance of $\delta
\approx 2\%$. The distribution above $\approx 350$ keV is described by a
trace contamination of the enriched and shield crystals by $^{40}$K, $^{137}$%
Cs, $^{113m}$Cd, two neutrino double beta decay of $^{116}$Cd with $%
T_{1/2}=2.6\times 10^{19}$ yr \cite{Dan00}, and external $\gamma $ rays.

The energy spectrum of $\alpha $ particles (Fig. 6) was obtained by
selection of events with the following values of shape indicator: $SI_\gamma
+4\sigma _\gamma <SI<SI_\alpha +2.4\sigma _\alpha $. Under such restrictions
the efficiency of the pulse-shape analysis ($\eta _{PSA}$) depends on the
energy of $\alpha $ particles. For example, for $\alpha $ peak of $^{180}$W
this efficiency equals 49.5\%, while additional use of the filter for the
plastic pulses discrimination decreases this value down to 47\%.

Taking into account the fact that secular equilibrium in crystal
scintillators is usually broken, the distribution of the $\alpha $ events is
well reproduced by the model (see insert in Fig.~6), which includes $\alpha $
decays from $^{232}$Th and $^{238}$U families. For illustration, the results
of the pulse-shape analysis of the data (for energy above 350 keV) are
presented in Fig. 7 as three-dimensional distribution of the background
events versus energy and shape indicator. In this plot one can see clearly
separated population of the $\alpha $ events, which belong to U/Th families.
The total $\alpha $ activity associated with the peak in the energy region $%
400-1500$ keV is 1.40(10) mBq/kg. However, because of a poor energy
resolution for $\alpha $ particles and uncertainty of the $\alpha $/$\beta $
ratio, we give only limits on the activity of nuclides from uranium and
thorium families in the $^{116}$CdWO$_4$ scintillators received from the fit
of the spectrum in the range $400-1500$ keV: $^{232}$Th $\leq $ 0.15 mBq/kg,
$^{238}$U ($^{234}$U) $\leq $ 0.6 mBq/kg, $^{230}$Th $\leq $ 0.5 mBq/kg, $%
^{210}$Pb $\leq $ 0.4 mBq/kg.

\nopagebreak
\begin{figure}[ht]
\begin{center}
\mbox{\epsfig{figure=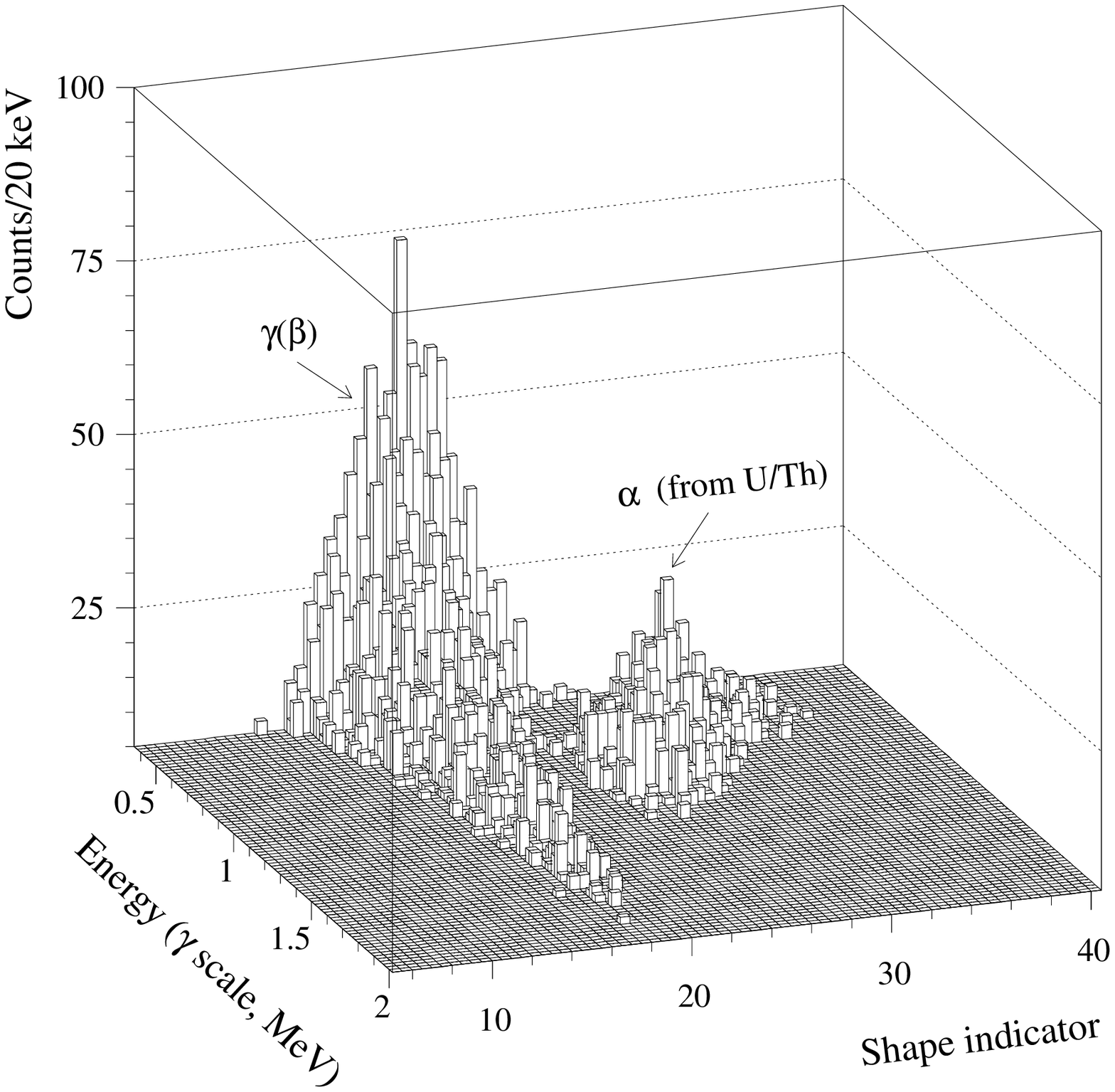,height=10.0cm}}
\caption{Three-dimensional distribution of the background events
(2975 h of exposition with the $^{116}$CdWO$_4$ crystals) versus energy and
shape indicator. The population of $\alpha $ events belonging to the U/Th
families is clearly separated from $\gamma (\beta )$ background.}
\end{center}
\end{figure}

The low energy part of the $\alpha $ spectrum (below 200 keV) can be
explained by the PMT noise, residual $\gamma (\beta )$ background, $\gamma $
or $\beta $ events in the $^{116}$CdWO$_4$ scintillators with small
admixture of the plastic light guide pulses (which were not discriminated by
the pulse-shape analysis), decays of Th and U daughter $\alpha $ nuclides
located inside defects or inclusions of the crystals (in that case alpha
particles can lose a part of their energy without scintillation light
emission), etc.

Lastly, a small peak is visible in the $\alpha $ spectrum of Fig. 6 at the
energy around 300 keV. Since the 2.46 MeV $\alpha $ peak of $^{180}$W is
expected at $307\pm 24$~keV (with FWHM = $110$ keV), and because the
position of $\alpha $ peak of $^{232}$Th (with the lowest $\alpha $ energy
of 4.0 MeV among all $\alpha $ emitters from U/Th families) must be at $%
\approx $600 keV, it is unlikely that this small peak can be attributed to
origin other than $\alpha $ decay of $^{180}$W. However, because a platinum
crucible was used for the growth of our crystals, we analyze the possible
imitation of the effect by $\alpha $ activity of $^{190}$Pt (abundance of $%
^{190}$Pt is 0.014\% \cite{abundance}, $T_{1/2}=6$.5$\times $1$0^{11}$ yr, $%
E_\alpha $ = 3164(15) keV \cite{Fir96} or $435\pm 30$ keV in $\gamma $
scale). Calculation shows that the detected peak can be caused by platinum
pollution (homogeneously spread in the volume of the $^{116}$CdWO$_4$
scintillators) at the level of $\approx 3$ ppm. To estimate the actual
platinum contamination in cadmium tungstate crystals, the results of
previous low background measurements performed by the Milano-Kiev
collaboration with the CdWO$_4$ crystal\footnote{%
This crystal was produced in the same apparatus as enriched ones.} of 58 g
\cite{Ale94} were considered. This experiment has been carried out in the
Gran Sasso Underground Laboratory, and CdWO$_4$ crystal was used as a low
temperature ($\approx $25 mK) bolometer. Energy resolution (FWHM) of the
detector was equal to 5 keV at 2.6 MeV. No events were registered in the
energy region of $3100-3300$ keV during 340 h, which allows us to set bound
on the $^{190}$Pt activity. In accordance with \cite{PDG00,Fel98} it yields
1.3 counts as limit for number of events, which can be excluded with 68\%
C.L., hence one can set a bound for platinum concentration in cadmium
tungstate crystal as 1.2 ppm. Besides, two samples of CdWO$_4$ crystal with
dimensions 1.5$\times $1.5$\times 0.$1 cm were searched (with the help of
the electron microscope) for inclusions, whose elemental composition is
different from that of CdWO$_4$. With this aim flat surface (2 cm$^2$) of
each sample was scanned, and if such an inclusion was observed, the electron
beam was concentrated on it and the emitted X rays were analyzed with a
crystal spectrometer (energy resolution is better than 0.1 eV) tuned on a
characteristic platinum X rays. For any observed inclusions (with diameters
in the range 2--30 $\mu $m) no emission of platinum X rays was found. It
results in conclusion that limit (95\% C.L.) on platinum concentration in
CdWO$_4$ crystal is lower than 0.1 ppm (or 0.03 ppm) for Pt inclusions with
diameters less than 5 $\mu $m (or 3 $\mu $m). Similar analysis performed
with the small sample ($\approx $3 cm in diameter) of the enriched $^{116}$%
CdWO$_4$ crystal gives even more stringent restrictions\footnote{%
Due to larger scanned area and better spatial resolution of the electron
microscope used in this case.}: 0.07 ppm (or 0.02 ppm).

Therefore, one can conclude that all bounds obtained for homogeneous Pt
contaminations in the CdWO$_4$ crystals are well below the level of 3 ppm,
at which $\alpha $ peak of $^{180}$W could be imitated. Moreover, it is also
known that in crystals, which were grown in platinum crucible, Pt is not
distributed homogeneously, but it is present in form of precipitates with
size around 20 $\mu $m \cite{Grab84}, hence a broad energy distribution
instead of an $\alpha $ peak would be observed. The latter was proved by our
Monte Carlo simulation of the $^{190}$Pt $\alpha $ decays in CdWO$_4$
crystal with the help of the GEANT and event generator DECAY4 \cite{Decay4}.
It gives that effect of $^{180}$W $\alpha $ activity could be imitated by $%
^{190}$Pt alpha decays only in the case of platinum particles of $2-3~\mu $m
size, with Pt average concentration in the crystals at the level of $4-6$
ppm, which is much larger than our experimental limits.

Nevertheless, because it is impossible to exclude (at least in principle)
some other explanations of the 300 keV peak in the background spectrum, we
can treat our experimental result only as the first indication for the
possible $\alpha $ decay of $^{180}$W. Obviously, final confirmation of its
existence would be obtained with the help of CdWO$_4$ crystals
enriched/depleted in $^{180}$W. However, it is clear that preparation and
performance of such a measurement would require a strong additional efforts
and perhaps a long time.

\subsection{Alpha activity of $^{180}$W and other tungsten isotopes}

Assuming that observed peak at the energy around 300 keV is really caused by
$\alpha $ decay of $^{180}$W, let us estimate its half-life. With this aim
the experimental $\alpha $ spectrum in the energy region of interest was
fitted by a Gaussian distribution (FWHM = $110$ keV), which represents the
effect, and by the background model. The latter was built up as a sum of the
$\alpha $ peak of $^{232}$Th and exponential function. Position and area of
the searched peak, the thorium $\alpha $ peak area, and constants of the
exponent were taken as free parameters of the fit procedure, which was
performed in the energy region (140$\pm $20) $-$ (510$\pm $30) keV. The best
fit ($\chi ^2$/n.d.f. = 0.52) achieved in the interval $140-520$ keV (see
Fig.~8) gives the area of the searched peak $S=$ 64$\pm $26 counts and its
position at 326$\pm $15 keV. Therefore, measured energy of $^{180}$W alpha
particles is $2580\pm 290$ keV, which is in a reasonable agreement with the
theoretical value $E_\alpha $=2460 keV. Taking into account number of $^{180}
$W nuclei in the crystals ($6.5$7$\times $1$0^{20}$) and the total
efficiency (47\%), we get the corresponding half-life value: $T_{1/2}^\alpha
(^{180}$W) = 1$.1_{-0.4}^{+0.8}$(stat)$\pm 0.3$(syst)$\times 10^{18}$ yr.
The systematic error is related mainly with an uncertainty of the background
model.

\nopagebreak
\begin{figure}[ht]
\begin{center}
\mbox{\epsfig{figure=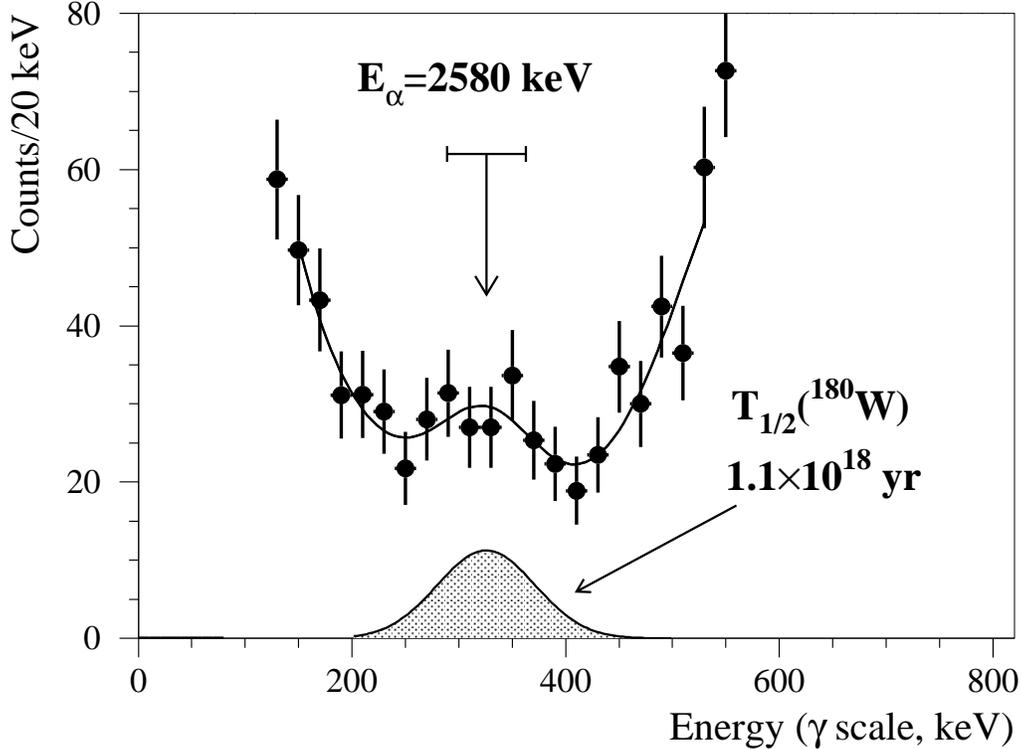,height=10.0cm}}
\caption{Fragment of the $\alpha $ spectrum measured with $^{116}$CdWO%
$_4$ detector during 2975 h together with the fitting curve (solid line).
The $\alpha $ peak of $^{180}$W with area of 64 counts corresponds to
half-life $1$.1$\times $1$0^{18}$ yr.}
\end{center}
\end{figure}

In more conservative approach we use results of our fit in order to set the
upper limit on half-life of $^{180}$W: lim $T_{1/2}^\alpha (^{180}$W) $\geq $
0.7$\times 10^{18}$ yr at 90\% C.L.\footnote{%
Similar bound $T_{1/2}^\alpha $ $\geq $ 1$\times 10^{18}$ yr at 90\% C.L.
can be extracted from the mentioned measurements (340~h) with the CdWO$_4$
crystal of 58 g used as a low temperature bolometer \cite{Ale94}, where no
events were observed within the energy interval 2516$\pm $30 keV.}

In addition, since there are no structural features in the experimental $%
\alpha $ spectrum, which could indicate an $\alpha $ activity of other
tungsten isotopes, half-life bounds for these processes were estimated. The
numbers of candidate nuclei in the detector with mass of 330 g are: $^{182}$%
W -- $1.4$5$\times $10$^{23}$, $^{183}$W -- $7.8$3$\times $1$0^{22}$, $%
^{184} $W -- $1.6$8$\times $1$0^{23}$, $^{186}$W -- $1.$56$\times $1$0^{23}$%
. The position of $\alpha $ peak in the $\gamma $ equivalent scale, the
expected peak width and the efficiency of the pulse-shape analysis are: $%
^{182}$W ($E_\alpha ^\gamma =215\pm $5$7$ keV; FWHM$_\alpha =86$ keV; $\eta
_{PSA}$ = 35\%), $^{183}$W ($E_\alpha ^\gamma =20$6$\pm 54$ keV; FWHM$%
_\alpha =83$ keV; $\eta _{PSA}=34\%$), $^{184}$W ($E_\alpha ^\gamma =$204$%
\pm 53$ keV; FWHM$_\alpha =83$ keV; $\eta _{PSA}=34\%$), $^{186}$W ($%
E_\alpha ^\gamma =14$7$\pm $34 keV; FWHM$_\alpha =69$ keV; $\eta _{PSA}=28\%$%
). Fitting the experimental $\alpha $ spectrum by sum of expected peak and
background model (exponential function plus $\alpha $ peaks of $^{180}$W and
$^{232}$Th) we get the following half-life limits (at 90\% C.L.):

\begin{center}
$T_{1/2}(^{182}$W$)\geq 1.$7$\times $1$0^{20}$ yr, \qquad \qquad $%
T_{1/2}(^{183}$W$)\geq 0.$8$\times $1$0^{20}$ yr,

$T_{1/2}(^{184}$W$)\geq 1.$8$\times $1$0^{20}$ yr, \qquad \qquad $%
T_{1/2}(^{186}$W$)\geq 1.$7$\times $1$0^{20}$ yr.
\end{center}

The obtained experimental results for $\alpha $ activity of naturally
occurring tungsten isotopes are summarized in Table.

\subsection{Comparison with theory}

To our knowledge, there is only one theoretical work, based on the
microscopical approach, in which the half-life for $\alpha $ decay of $^{180}
$W was calculated \cite{Alb88}. It takes into account the systematic
behaviour of the reduced $\alpha $ widths of even-even nuclei with numbers
of neutrons and protons $N\geq 84$, $Z\leq 84$ and penetration factors
obtained from realistic cluster wave functions. The derived result is: $%
T_{1/2}^\alpha $ = $7.$5$\times $1$0^{17}$ yr and uncertainty of this value
is estimated as less than a factor of 3 \cite{Alb88}. For other tungsten
isotopes theoretical predictions are absent.

We have calculated the half-lives of all W isotopes for $\alpha $ decay,
first, with the help of the cluster model \cite{Buc91}, which was very
successful in describing the $\alpha $ decays of even-even nuclei. For
example, it reproduces the $T_{1/2}^\alpha $ values mainly within a factor
of 2 for a wide range of nuclides (from $_{~52}^{106}$Te to $_{108}^{264}$%
Hs) and for $T_{1/2}^\alpha $ from $10^{-7}$ s to $10^{16}$ yr. Because $%
\alpha $ decays of natural W isotopes (except $^{183}$W) occur without
changes in nuclear spin and parity, we have chosen a set of 17 nuclides
(from $_{~60}^{144}$Nd to $_{~96}^{248}$Cm) with $T_{1/2}^\alpha >10^5$ yr
\cite{Nud00}, whose $\alpha $ decays also proceed without changes in nuclear
spin and parity. While comparing a theory with an experiment, it is
practical to use the value of the deviation $\chi =\max (R,1/R)$, where $R=$
$T_{1/2}^{th}/T_{1/2}^{exp}$. For the chosen set of nuclides the cluster
model \cite{Buc91} gives a quite reasonable average value of $\chi =1.$9.
Then, half-lives for W isotopes were calculated with the help of the cluster
model \cite{Buc91} and obtained results are listed in Table (given
uncertainties are related only with uncertainties of $Q_\alpha $). In
particular, $T_{1/2}^\alpha (^{180}$W) $=$ 8.3$\times $1$0^{17}$ yr, that is
very close to the value of $7.$5$\times $1$0^{17}$ yr derived in
\cite{Alb88}.

Semiempirical relationships are often more effective in the prediction of $%
T_{1/2}^\alpha $ than microscopically based calculations. We found in the
literature \cite{Poe83,Poe86,Fro57,Wap59,Taa61,Vio66,Kel72,Hor74,Bro92,Mol97}
18 semiempirical formulae which can be used for our purpose. All of them
were also tested with the same set of 17 nuclides with $T_{1/2}^\alpha >10^5$
yr. The best result (average deviation of the calculated values from
experiment $\chi =1.9$) was found for the relationship of \cite{Poe83} based
on phenomenological fission theory of $\alpha $ decay and valid not only for
even-even but also for odd-even, even-odd and odd-odd nuclei. The values of $%
T_{1/2}^\alpha $ obtained in such approach for W isotopes are given in
Table, where uncertainties are caused, as in the previous case, by
uncertainties of $Q_\alpha $. For the $^{183}$W decay it was also taken into
account that change in parity will additionally suppress the decay rate as
compared with that of \cite{Poe83}. For $^{180}$W our result $T_{1/2}^\alpha
=2$.0$\times $1$0^{18}$ yr is also close to that of \cite{Alb88}.

Thus, we can conclude that half-life value ($T_{1/2}^\alpha $ = 1$.$1$%
_{-0.5}^{+0.9}\times $1$0^{18}$ yr) for possible $\alpha $ decay of $^{180}$%
W (indication for which is observed in present work) is in a good agreement
with the theoretical predictions: $T_{1/2}^\alpha =0.$75$\times $1$0^{18}$
yr (microscopic approach \cite{Alb88}), and calculated here on the basis of
semiempirical formula \cite{Poe83} and cluster model \cite{Buc91} as $%
T_{1/2}^\alpha =2.$0$\times $1$0^{18}$ yr and $T_{1/2}^\alpha =0.8$3$\times $%
1$0^{18}$ yr, correspondingly.

\section{CONCLUSIONS}

In the present work the pulse shape and $\alpha /\beta $ ratio of the CdWO$%
_4 $ crystal scintillators have been studied for three directions of the
collimated beam of $\alpha $ particles relatively to the main crystal axes
in the energy range of $0.5-5.2$ MeV. The dependences of $\alpha /\beta $
ratio and pulse shape on the direction of $\alpha $ irradiation have been
found and measured.

By using the super-low background $^{116}$CdWO$_4$ crystal scintillators,
the indication for the $\alpha $ decay of natural tungsten isotope $^{180}$W
was observed for the first time. The measured half-life $T_{1/2}^\alpha $~=~1%
$.1_{-0.4}^{+0.8}$(stat)$\pm 0.3$(syst)$\times 10^{18}$ yr is close to the
theoretical predictions, and it would be the most rare $\alpha $ decay ever
observed in nature. More conservatively our result can be treated as the
lower limit on half-life of $^{180}$W: lim $T_{1/2}^\alpha $($^{180}$W) $%
\geq $ 0.7$\times 10^{18}$ yr at 90\% C.L. Besides, new $T_{1/2}$ bounds
have been set for $\alpha $ decay of $^{182}$W, $^{183}$W, $^{184}$W and $%
^{186}$W at the level of $(0.8-1.8$)$\times $1$0^{20}$ yr. All these limits
are higher than those obtained in previous work \cite{Geo95} and are the
most stringent bounds on $T_{1/2}^\alpha $ for any $\alpha $ decaying
nuclides.

To confirm existence of $\alpha $ activity of $^{180}$W, we are preparing
measurements with other tungsten containing crystal scintillators: CdWO$_4$
(whose scintillation characteristics are better than those of currently used
crystals), CaWO$_4$, and ZnWO$_4$. Observation of the $\alpha $ decay of $%
^{180}$W could be also checked with these crystals as bolometers \cite
{Ale94,Sis00} and, apparently, sensitivity of such experiments would be
enhanced by using crystals enriched/depleted in $^{180}$W.

\section{Acknowledgments}

The authors would like to thank Dr. Filippo Olmi for the electron microscope
measurements of the platinum contaminations in the CdWO$_4$ crystal and Dr.
Mykola Petrenko for the similar measurements performed with the enriched $%
^{116}$CdWO$_4$ crystal.

\end{document}